\documentclass[prb,twocolumn,aps,superscriptaddress]{revtex4}
\usepackage{graphicx}
\usepackage{subfigure}

\begin{document}

\title{Quantum Monte-Carlo study of a two-species boson Hubbard model}
\author{Siegfried Guertler}
\address{Center of Theoretical and Computational Physics and Department of Physics, The University of Hong Kong, Hong Kong, China}
\author{Matthias Troyer}
\address{Theoretische Physik, ETH Z\"urich, 8093 Z\"urich, Switzerland}
\author{Fu-Chun Zhang}
\address{Center of Theoretical and Computational Physics and Department of Physics, The University of Hong Kong, Hong Kong, China}
\date{\today }

\begin{abstract}
We consider a two-species hard-core boson Hubbard model for a supersolid,
where the two types of bosons represent vacancies and interstitials doped into a commensurate crystal. 
The on-site inter-species interaction may create bound states of vacancies and interstitials 
facilitating vacancy condensation at lower energies than in a single-species model, as suggested in an 
earlier mean field study. Here we carry out quantum Monte Carlo simulation to study possible supersolid 
phases of the model, corresponding to superfluid phases of the vacancies or interstitials. 
At low temperatures, we find three distinct superfluid phases. The extent of the phases 
and the nature of the phase transitions are discussed in comparison to mean-field theory.
\end{abstract}

\maketitle

\section{Introduction}

A supersolid is a special type of solid with superfluid properties. It has a diagonal particle density long range order
as in a usual crystal, and an off-diagonal long range order in particle density as in a superfluid.  
The simplest model for supersolid was proposed by Andreev and Lifshitz in 1969.\cite{SS1}
Their model was introduced to describe possible supersolid phase in Helium-4. 
In their model, vacancies or interstitials of solid Helium may 
exist in the ground state and condense due to the large quantum fluctuation of Helium atoms.  
The interaction between vacancy and interstital is neglected in their model. 

In this paper we study a two-species boson Hubbard model, which is 
an extension of the Andreev-Lifshitz model to include the interaction between vacancy and interstitial. 
This two-species model was recently introduced by Dai, Ma, and Zhang,\cite{PRBDMF} motivated by the observation 
of non-classical rotational inertia moment in solid helium-4 reported by Kim and Chan.\cite{KCNATSCI}
They used a mean field theory to study the ground state of the model and the possibility of the supersolid phase.
It was shown that the interaction of vacancies and interstitials may facilitate a supersolid phase. In this paper we use 
quantum Monte Carlo (QMC) simulations to study the possible supersolid 
and the finite temperature phase transition in the two-species boson model. 
The simulations support the qualitative conclusion obtained in the mean field theory
that the vacancy-interstitial interaction may facilitate supersolidity. 
Using QMC, we calculate the phase diagram, the superfluid densities of bosons, and 
the specific heat of the system. The two-species boson model and our calculations may be useful 
to understand other boson problems such as bosons in 
optical lattices.\cite{Zhuang}

Before we present the model and our results, we briefly summarize the current situation in study of supersolid Helium-4.
Because of its light mass and its bosonic nature, solid helium-4 has been a natural candidate for possible
supersolid at low temperatures and high pressures. Theoretically, such a possibility 
was proposed by Andreev and Lifshitz\cite{SS1} and by Chester. \cite{SS2}
Leggett further predicted the non-classical rotational inertia moment of such a supersolid in a rotating 
experiment.\cite{SS3,SS4} The interest of supersolid has been recently revived 
due to the observation of non-classical rotational inertia 
in solid helium-4.\cite{KCNATSCI}
By now, the non-classical inertia moment in solid helium has been confirmed by other groups.\cite{Reppy,Kondo,Penzev}
However, it remains controversial 
if the phenomenon is related to the supersolidity and if the supersolid phase is a bulk equilibrium phenomenon.
\cite{Beamis,Sasaki,chan2} On the theoretical side QMC simulations did not find a supersolid phase in 
Helium-4.\cite{Ceperley,Boninsegni,FPC1}
Furthermore, the vacancies or interstitials in helium are shown 
to attract to each other and to tend to have phase-separation,\cite{FPC1} indicating that the 
Andreev-Lifshitz model may not describe solid helium.

\section{Model and Method}

We consider a two-species boson Hubbard model in a cubic lattice with $z=6$ nearest neighbors:
\begin{eqnarray}
H=\sum_{j} (\epsilon_a n_{j,a} + \epsilon_b n_{j,b} - U n_{j,a} n_{j,b}) -  \\ \nonumber
\sum_{\langle i,j \rangle}( t_{a} a^{\dagger}_i a_j +
 t_{b} b^{\dagger}_i b_j + h.c.) 
\end{eqnarray}
where $a_j$ is an annihilation operator of boson $a$ at lattice site $j$, representing a vacancy, and 
$b_j$ an annihilation operator of boson representing an interstitial, in a vacuum representing a defect-free insulating crystal
of bosonic atoms. 
$n_{j,a}=a^{\dagger}_j a_j$ and $n_{j,b}=b^{\dagger}_j b_j$ are the number operators 
for $a$- and $b$-bosons, and $\epsilon_a$ and $\epsilon_b$ are site boson energies, respectively. 
We consider the interesting case 
$\epsilon_a>0$, and $\epsilon_b>0$. We assume both vacancy and interstital are hard-core bosons, so that the allowed
values for $n_{j,a}$ and $n_{j,b}$ are either 0 or 1. An exciton is described by the state with both a vacancy and an interstitial at the same lattice site $n_{j,a}=n_{j,b}=1$.
The couplings $t_a$ and $t_b$ are the hopping integrals for boson
$a$ and $b$, respectively, and we assume $t_a$ and $t_b$ to be positive without loss of generality. 
$U$ is the on-site attractive interaction between a vacancy and an interstitial. Note that
the attractive interaction between a vacancy and an interstitial reflects the strong short range 
repulsion between two nearby atoms when an interstitial atom is added into the lattice. 

Without interaction, at $U=0$, the two-species model decouples into independent vacancy and interstitial models.
The ground state of the $a$-boson (vacancy) model  is superfluid (a vacancy supersolid) if $zt_a > \epsilon_a$ 
and an empty vacuum state (insulating solid) otherwise.
Similarly, the ground state of the $b$-boson (interstitial) model  is superfluid  (an interstitial supersolid)
if $zt_b > \epsilon_b$ and the empty vacuum state (an insulating solid) otherwise.

The attractive inter-species boson interaction $U$ couples the two types of boson,
and the problem cannot be solved analytically
without approximation.  This model was studied by using a mean field 
theory at zero temperature,\cite{PRBDMF} and a special limiting case with 
$\epsilon_b \rightarrow \infty$ but a finite  $ \epsilon_b-U$  was investigated by a modified spin wave 
theory.\cite{Zhuang} The main effect of the attractive term is to facilitate vacancy or/and 
interstital condensation due to excitons or bound states. 

In this paper we will use QMC methods to study the phase transition and superfluid properties of the model. We use a slightly extended version of the directed loop algorithms\cite{DIRL1,DIRL2} of the ALPS project.\cite{ALPS} In the stochastic series expansion (SSE) representation used by this algorithm the superfulid 
density can be measured through the fluctuations of the spatial winding numbers.\cite{CeperlyPollock}
In three dimensions for a simple cubic lattice the relationship is:
\begin{equation}
\rho_{s,\alpha} = \frac{T}{L}  \langle W_{\alpha}^2 \rangle,
\end{equation}
where $\alpha=a,b$ refers to the type of boson, $W_\alpha$ is the spatial winding number of the bosons in 
one direction, $L$ is the linear size of the cubic lattice and $T$ the temperature.
In addition we consider the correlated winding numbers
\begin{equation}
W_{\pm} = \langle (W_{a} \pm W_{b})^2 \rangle = \langle W_{a}^2 \rangle + \langle W_{b}^2 \rangle + 2 \langle W_{a} W_{b} \rangle
\end{equation}
and define
\begin{equation}
\rho_{\pm}=\frac{T}{L}W_{\pm}.
\end{equation}

We performed simulations on lattices with up to $10^3$ lattice sites. 
Larger lattices did not equilibrate using the directed loop algorithm due to the formation of bound 
states between $a$ and $b$ bosons. A two-worm algorithm such as developed for the one-dimensional 
case in Ref. \onlinecite{Pollet} would be required to go to larger lattices, 
however we found that the sizes used here were sufficient 
to determine the nature of the phases.

\section{Symmetric case}

\begin{figure}[tbp]
     \centering
    \includegraphics[width=\columnwidth]{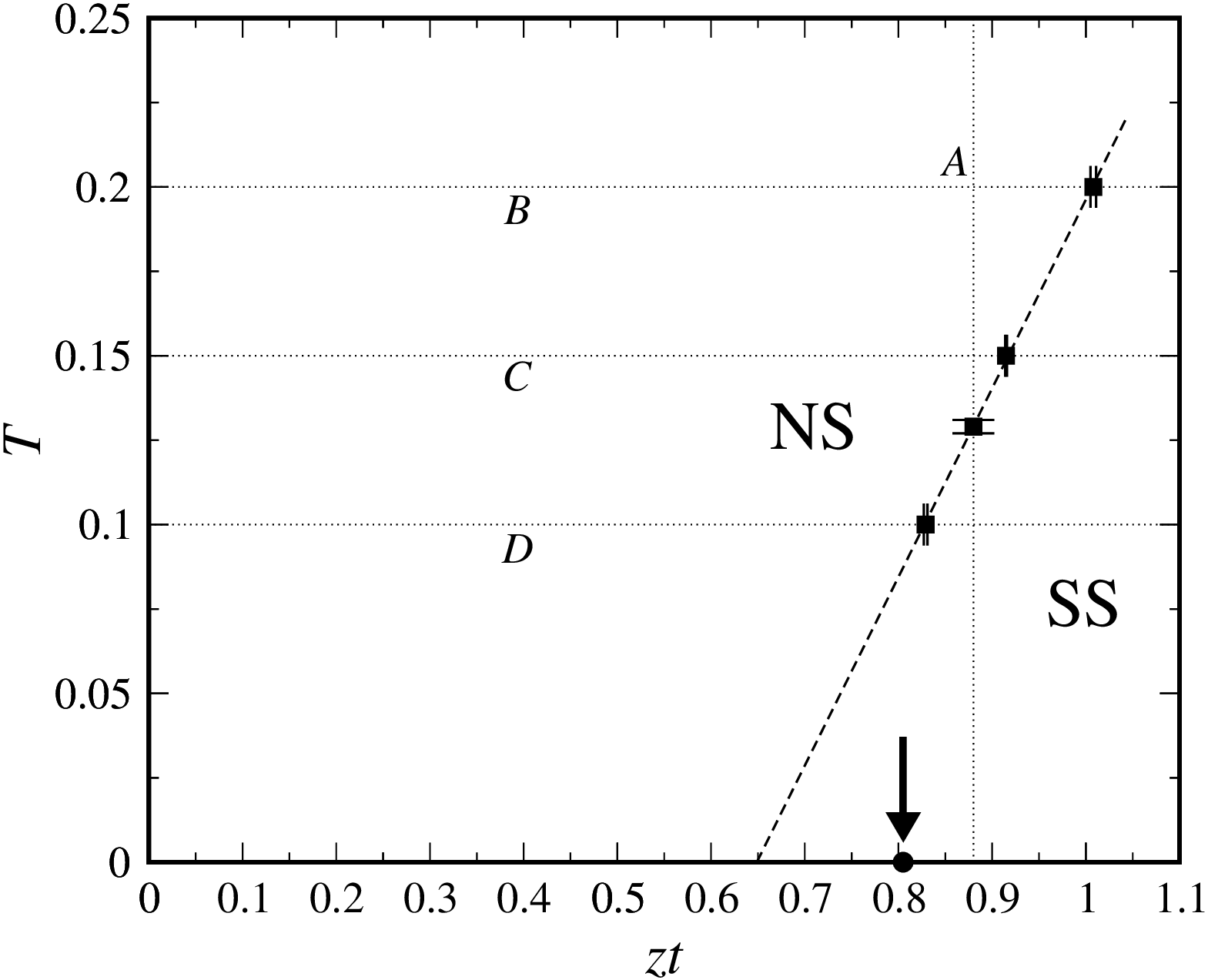}
    \caption{Finite temperature $T$ phase diagram of model (1) in the symmetric case $t_a=t_b=t$,
obtained by quantum Monte Carlo (squares with error bars). $\epsilon_a=\epsilon_b=1$, $U=1.8$. The dashed line is a linear fit 
to the data, separating supersolid phase (SS) from normal solid (NS). 
The mean field transition point at $T=0$ is indicated by a vertical arrow.\cite{PRBDMF}
Lines $A$ - $D$ indicate the parameters of cuts through the phase diagram that we will investigate in more detail in Figs.  \ref{VISFsym} and \ref{QPTSYM}.}
    \label{Tc_estimate_sym}
\end{figure}

We first consider the symmetric case with $\epsilon_a=\epsilon_b=1$. 
The energy cost of an exciton, (both an $a$- and a $b$-boson on the same lattice site), is
$\Delta= \epsilon_a+\epsilon_b-U$.  The system becomes a trivial exciton lattice if $\Delta <0$, which we will not discuss.
For the symmetric case,  we choose $U=1.8$ in our simulations to examine correlation effects.
In Fig. \ref{Tc_estimate_sym}, we summarize our result by showing the phase diagram in the parameter space of temperature $T$ and
boson hopping $zt=zt_a=zt_b$. The simulations are carried out at temperatures ranging from $0.1$ to $0.2$, 
which allow us to estimate a zero temperature phase boundary at $zt \approx 0.65$, 
smaller than the mean field value of 
$zt \approx 0.80$, and smaller than that of the non-interacting case of $U=0$ at $zt=1$.  Hence, quantum fluctuations which are neglected in the mean field theory 
further favor the superfluid phase. 

\begin{figure}[tbp]
     \centering
    \includegraphics[width=\columnwidth]{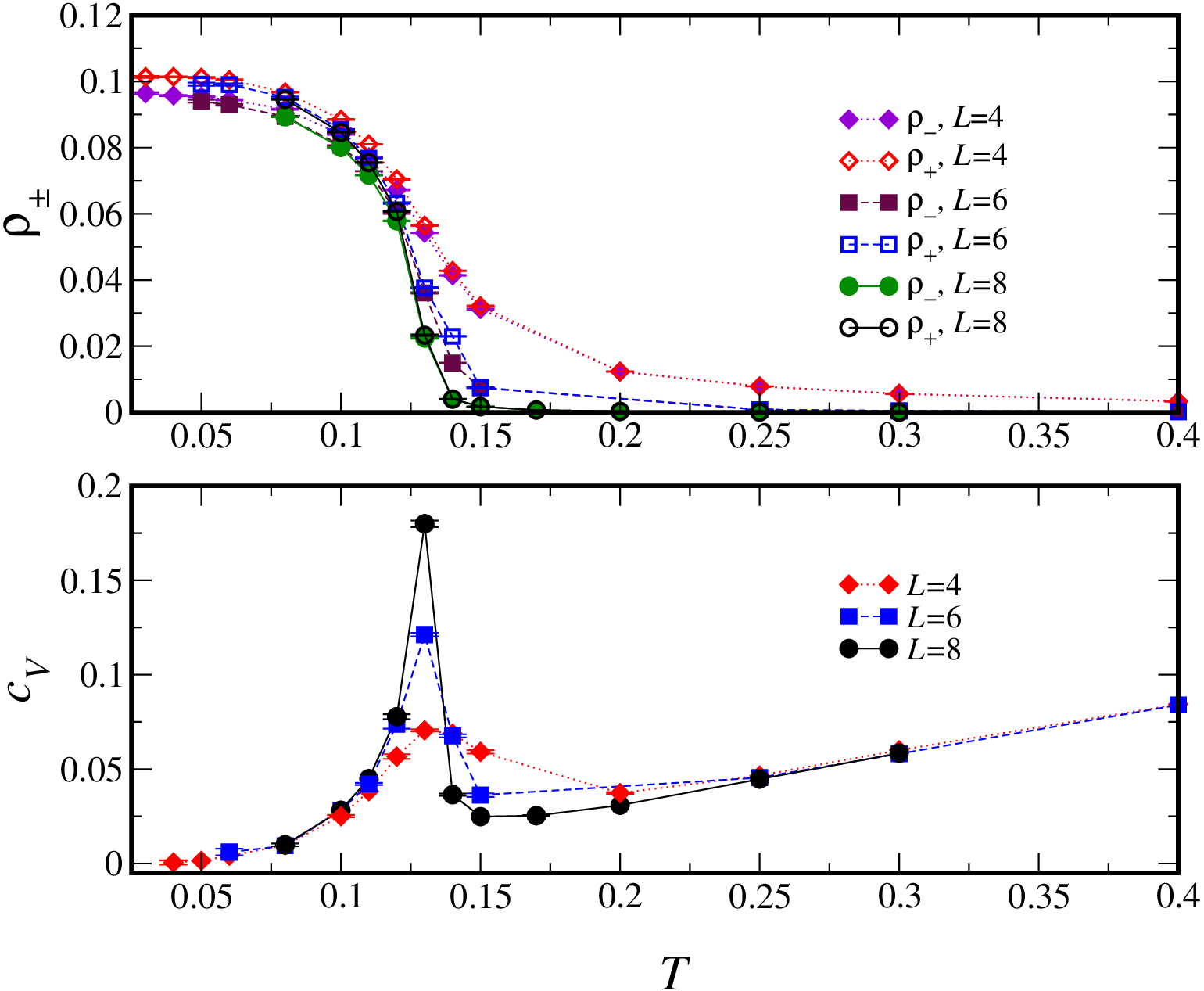}
    \caption{Superfluid densities $\rho_{\pm}$ and specific heat $c_V$ as functions of the temperature $T$,
at $zt_a=zt_b=0.88$ along line $A$ in Fig. \ref{Tc_estimate_sym}. 
}
    \label{VISFsym}
\end{figure} 

We examine the temperature dependences in more detail along line $A$ in the phase diagram of Fig. \ref{Tc_estimate_sym}. 
Figure \ref{VISFsym} shows the superfluid densities  $\rho_{\pm}$ and the specific heat $c_V$ as functions of $T$ along line $A$ in Fig. \ref{Tc_estimate_sym}
 ($zt=0.88$) at the temperature region from 
$T=0.03$ to $T=0.4$. As $T$ decreases, $\rho_{\pm}$ rise abruptly 
below $T=0.13$ and saturate to $\rho_{\pm} =0.1$,
$c_V$ develops a clear peak around $T=0.13$, and the peak becomes sharper as the size increases. 
Note that $\rho_+= \rho_-$ within our error bars, indicating that there are no correlations and the 
two types of bosons condense independently with the same superfluid density $\rho_a=\rho_b \approx \rho_+/2$.

\begin{figure}[tbp]
     \centering
    \includegraphics[width=\columnwidth]{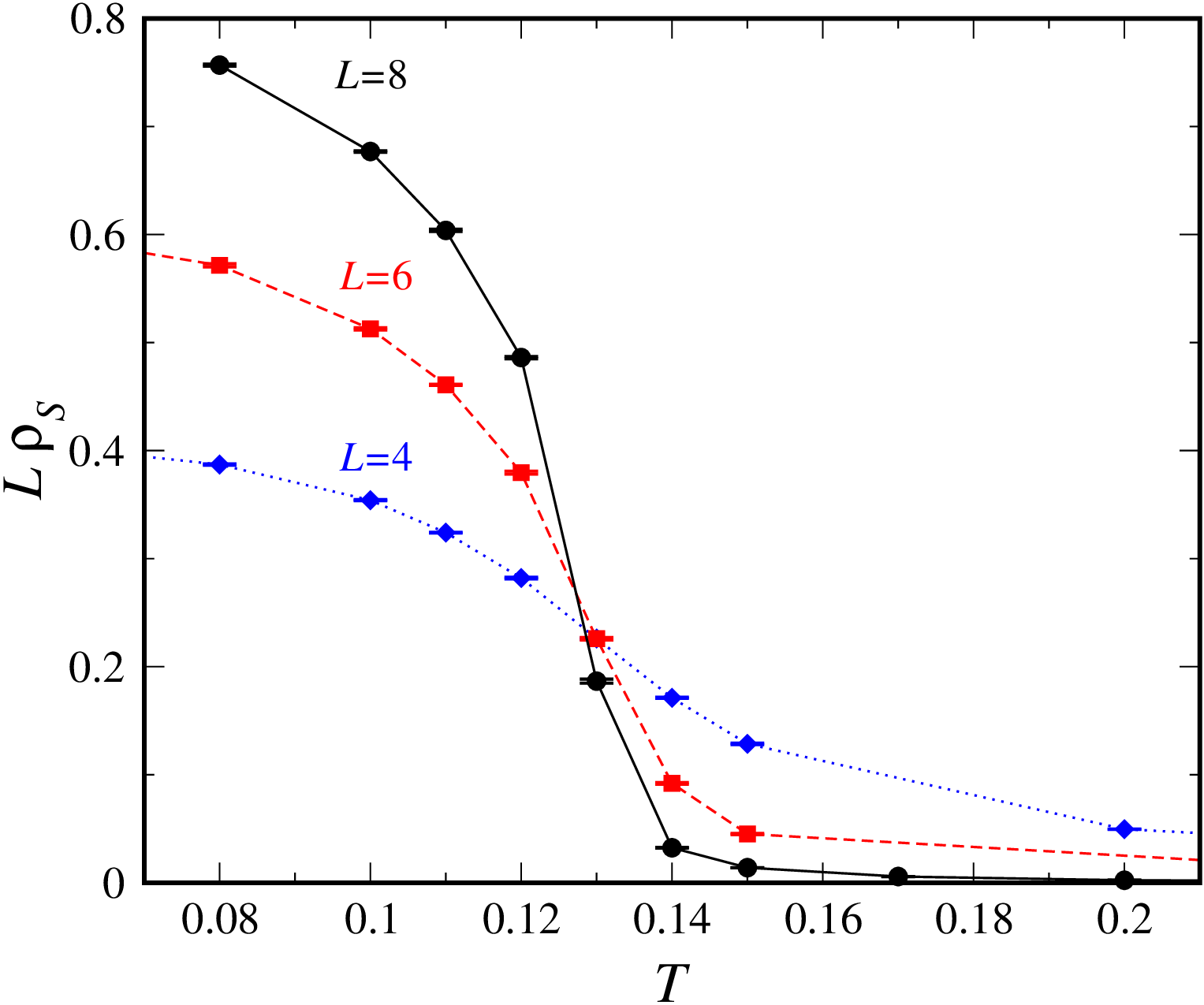}
    \caption{Superfluid density $\rho_s L$ as a function of $T$ for the symmetry model 
at $zt=0.88$, along line $A$ in Fig.\ref{Tc_estimate_sym}, for $L=4$, $6$ and $8$.  The three curves cross at one point, 
>from which we estimate $T_c =0.129 \pm 0.002$.}
    \label{VISFSYN_cross}
\end{figure}

In Fig. \ref{VISFSYN_cross}, we show the the superfluid density for system sizes $L=4$, $6$, $8$. 
Finite size scaling for a second order phase transition in the $U(1)$ universality class implies that 
$\rho_sL$ is a constant at the transition temperature $T_c$. The three curves in Fig. \ref{VISFSYN_cross} 
indeed cross at a single point, from which we can estimate $T_c =0.129 \pm 0.002$.  

\begin{figure}[tbp]
     \centering
    \includegraphics[width=\columnwidth]{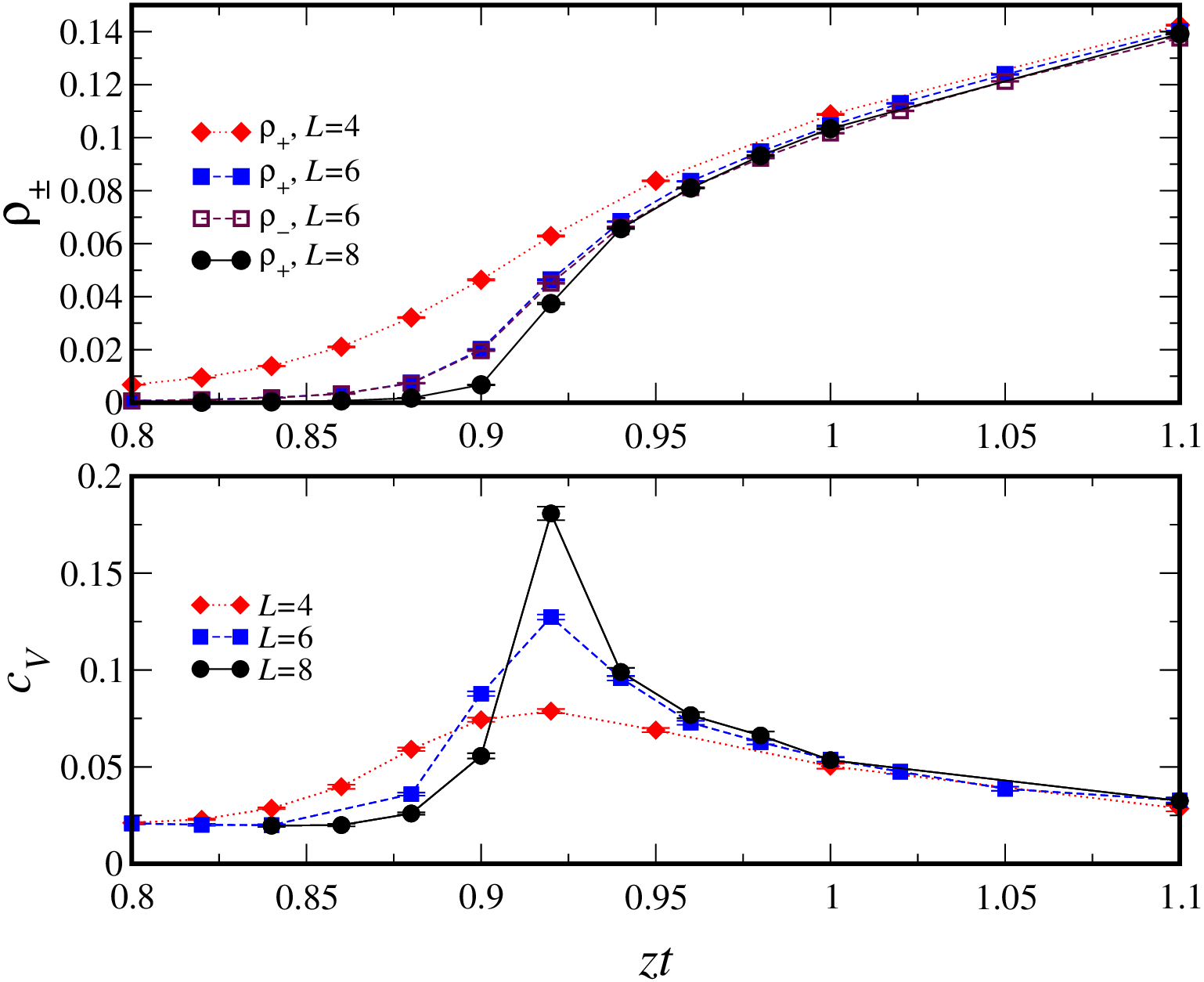}
    \caption{Superfluid densities $\rho_{\pm}$  and specific heat as functions of $zt$ in the symmetric model 
along line $B$ in Fig. \ref{Tc_estimate_sym}.
}
    \label{QPTSYM}
\end{figure} 

\begin{figure}[tbp]
     \centering
    \includegraphics[width=\columnwidth]{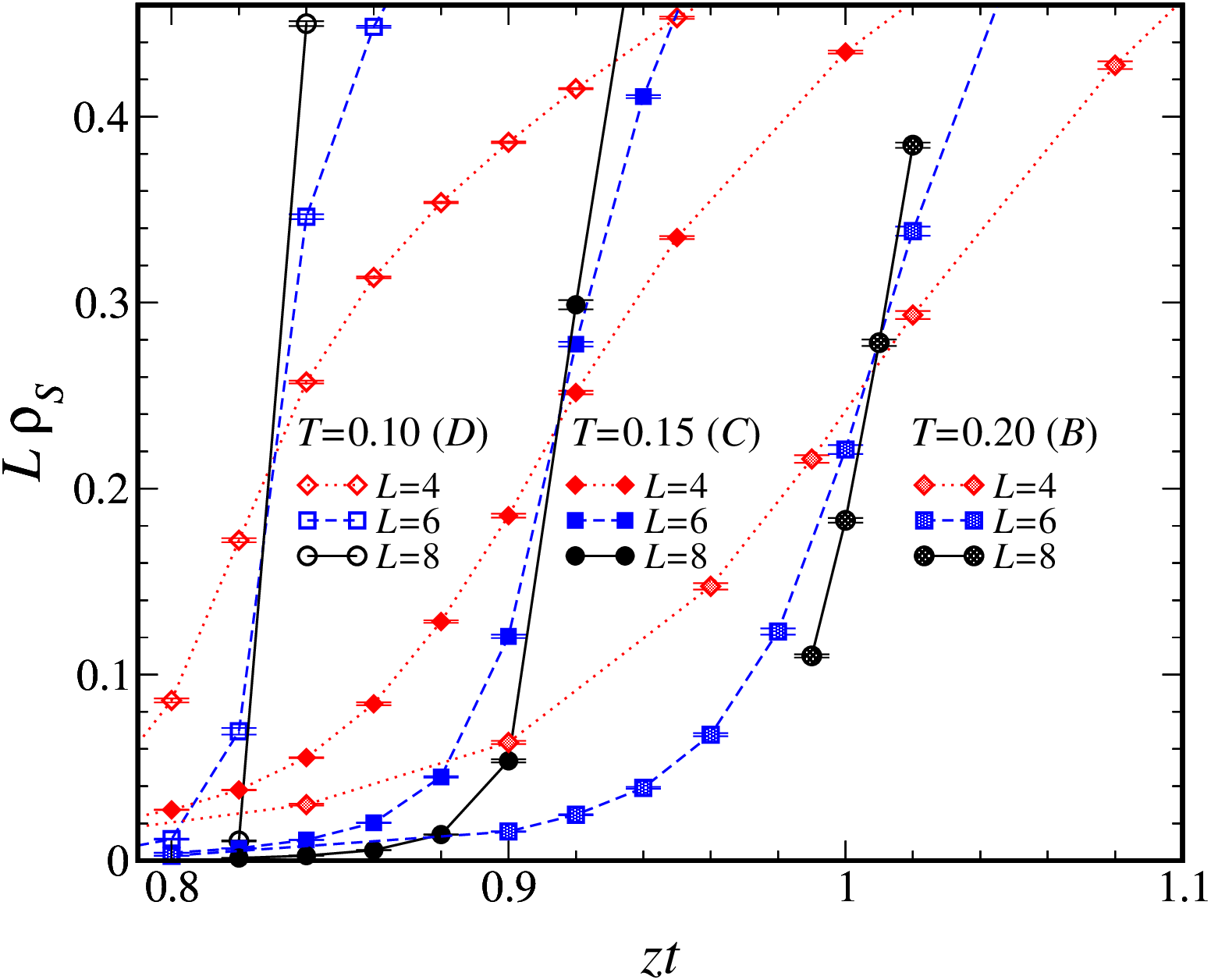}
    \caption{$\rho_s L$ as functions of $zt$ in the symmetric model 
at $T=0.1$, $0.15$, $0.20$ corresponding to lines $B$, $C$ and $D$ shown in Fig. \ref{Tc_estimate_sym}. 
For each temperature the curves for different system sizes cross at one point, consistent with a second order phase transition.
}
    \label{VISF_Wcross}
\end{figure}

Finally we investigate the dependence on the hopping amplitude $zt$ at various temperatures. 
In Fig. \ref{QPTSYM} we plot  the superfluid densities and specific heat at $T=0.15$ (along line $C$).  
Superfluidity develops at around $zt=0.92$ as we can see from both $\rho_+$ and $c_V$. 
The superfluid density as functions of $zt$ 
are plotted in Fig. \ref{VISF_Wcross} for different system sizes along lines $B$, $C$, and $D$. 
Each set of curves cross at one point, consistent with the expected scaling at a second order 
phase transition.

\section{Non-symmetric case}

We now discuss the non-symmetric case, which is more interesting and possibly more 
relevant to physical systems since there is a lack of vacancy-interstitial symmetry. 
In all the simulations reported for the non-symmetric model, we consider $\epsilon_a=1, \epsilon_b=4$, and $U=4$. 
We use smaller values of $U$ than in the mean-field work of Ref. \onlinecite{PRBDMF}  
since larger values of $U$ cost too much CPU time. 

\begin{figure}[tbp]
\begin{minipage}[t]{.23\textwidth}
\includegraphics[height=5cm]{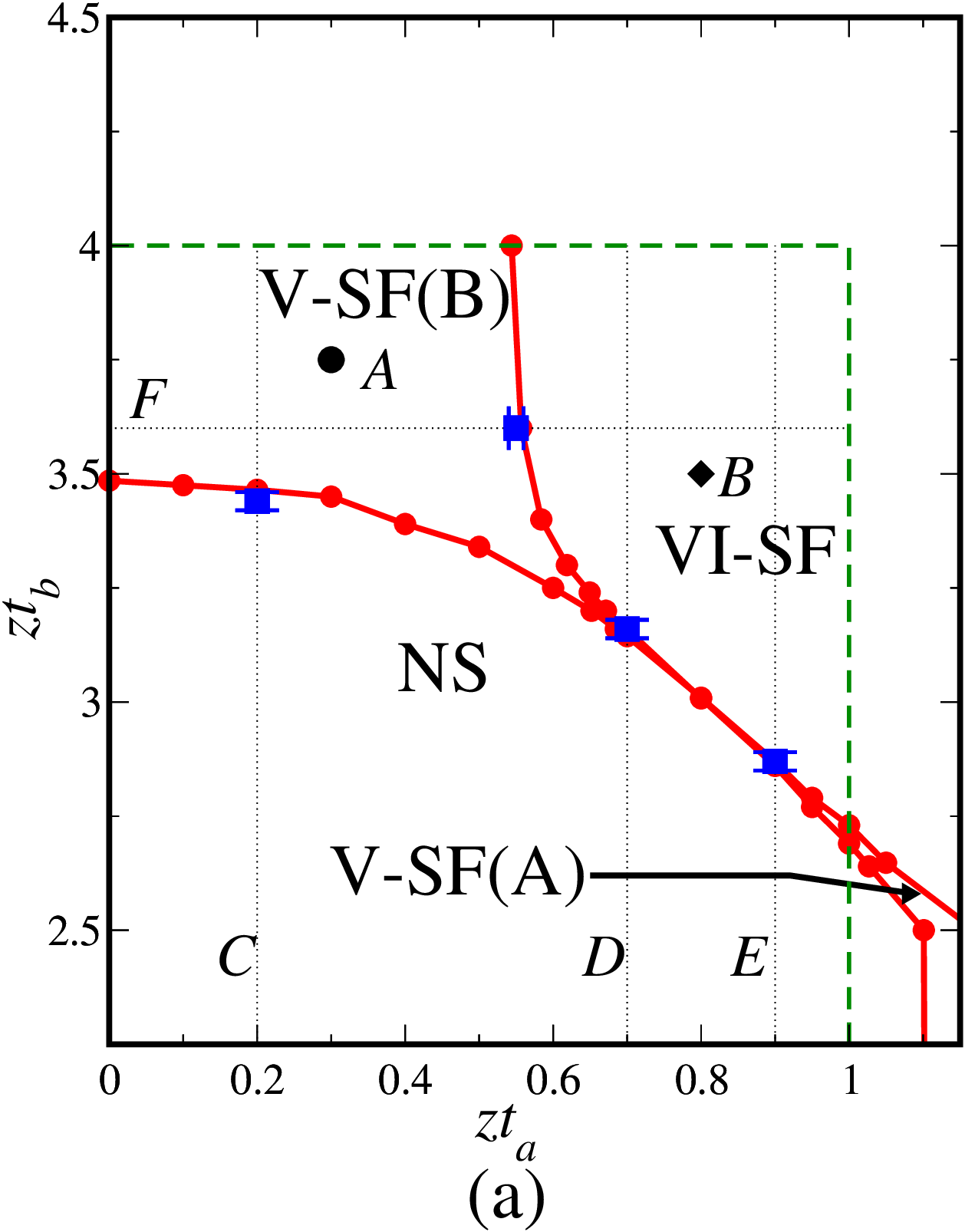}
\end{minipage}%
\begin{minipage}[t]{.23\textwidth}
\includegraphics[height=5cm]{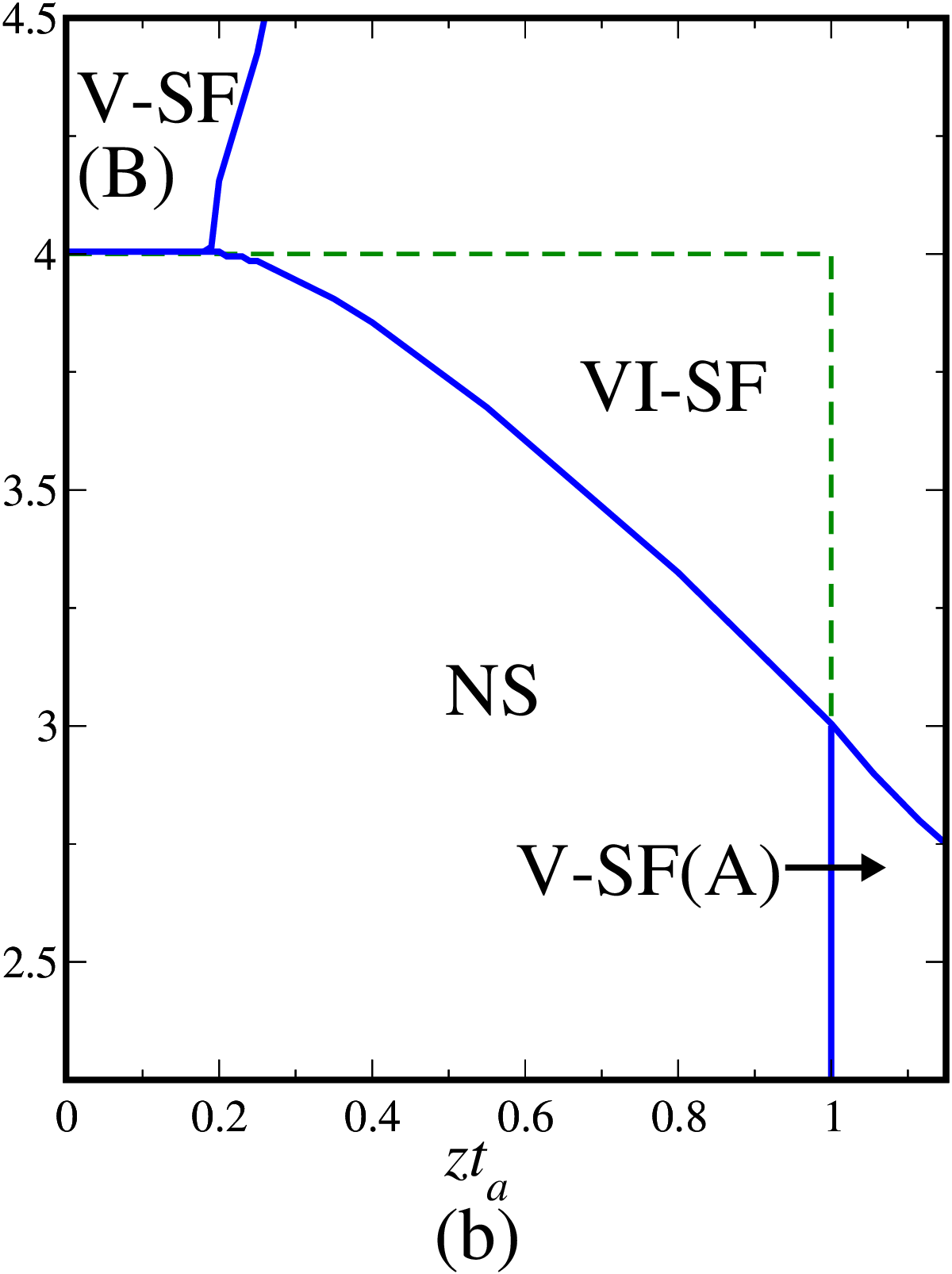}
\end{minipage}
\caption{
(a). Phase diagram obtained in QMC at $T=0.15$ for the non-symmetric case of model (1)
with parameters $U=4.0$, $\epsilon_a=1$ and $\epsilon_b=4$.  
The phase transition points marked by red circles are obtained from system sizes $L=3$ and $4$.
Blue squares are obtained from systems with $L=4$, $5$ and $6$.
Points $A$ and $B$ and the Lines $C$, $D$, $E$, $F$ indicate 
cuts through the phase diagram that we study in more detail in Figs. \ref{VSFB}-\ref{QPTztb36}.  
(b). Mean field ground state phase diagram for the same 
parameters, obtained using the mean field theory of Dai {\it et al}.\cite{PRBDMF}
Dashed green lines are the transition lines of the noninteracting model 
with $U=0$. 
}
\label{444T015phase}
\end{figure}

Our phase diagram
in the parameter space of $zt_a$ and $zt_b$ at $T=0.15$, is summarized in Fig. \ref{444T015phase}(a). 
This temperature is low enough to observe  
the expected supersolid phases. In addition to the normal solid, there are three
supersolid phases: 
\begin{enumerate}
\item a vacancy superfluid-A phase [V-SF(A)]  in which
the $a$-bosons (vacancies) condense $\rho_a\ne0$ and no $b$-bosons (interstitials) are present. \footnote{This phase is similar to the vacancy state of 
Andreev and Lifshitz.\cite{SS1}} 
\item a vacancy superfluid-B phase [V-SF(B)] in which the $b$-bosons condense $\rho_b\ne0$ and $n_{i,a}=1$ (for $T=0$). 
This is a vacancy superfluid above a background of excitons.
Vacancies move in an otherwise excitonic lattice, so it may be called vacancy superfluid.\cite{PRBDMF}
\item a vacancy and interstitial superfluid [VI-SF] phase in which both $a$- and $b$-bosons condense: $\rho_a\ne0$ and $\rho_b\ne0$.
\end{enumerate}


The phase boundaries labeled by red circles in Fig.~\ref{444T015phase}(a) are obtained 
from simulations on systems with up to $L= 4$.
Calculations on larger size systems in these parameter region require much more computational effort,
and are only carried out for four
selected points, labeled by blue squares on the boundaries in the figure,
representing typical interesting cases
of the three most interesting different phase-transitions in the parameter space. 

For comparison, we show in Fig. \ref{444T015phase}(b) 
the result of mean field calculations for the same parameters considered.
Note that the QMC predicts a larger parameter space for the supersolid phases than the mean field theory,
indicating again that the quantum fluctuation neglected in the mean field theory but included in the QMC 
is in favor of the supersolid phase.  

In the remaining part of this section, we discuss the phase transition as a function of temperature 
and as a function of boson hopping integrals.

\begin{figure}[tbp]
\includegraphics[width=\columnwidth]{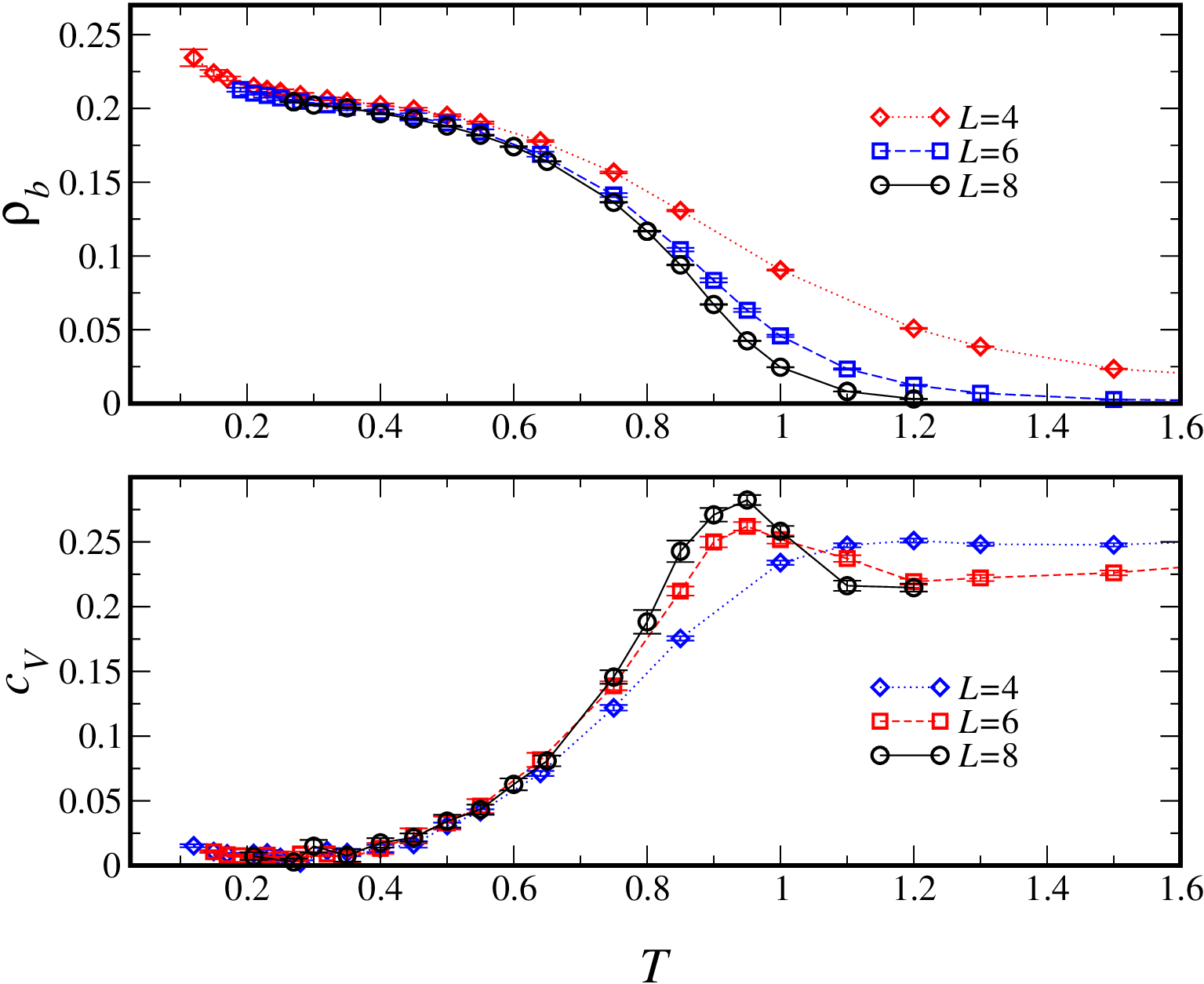}
\caption{Superfluid density $\rho_b$ of $b$-boson and specific heat as functions of $T$
for non-symmetric case at point $A$ in Fig. 6(a).  Here the low-temperature phase is V-SF(B).
}
\label{VSFB}
\end{figure}

To study the temperature dependence, we choose two typical 
points $A$ 
($zt_a=0.3$ and $zt_b=3.75$) and $B$ ($zt_a=0.8$ and $zt_b=3.5$)
in the parameter space as indicated in Fig. \ref{444T015phase}(a).
In Fig. \ref{VSFB}, we show the superfluid density and specific heat as functions of the temperature
for the system at point $A$. As the temperature decreases, $\rho_b$ starts to increase sharply 
at around $T\approx0.92$, while $\rho_a$ remains zero. This indicates that only $b-$bosons condense. 
A scaling analysis of $\rho_{b}L$ gives $T_c=0.924\pm0.002$.

\begin{figure}[tbp]
\includegraphics[width=\columnwidth]{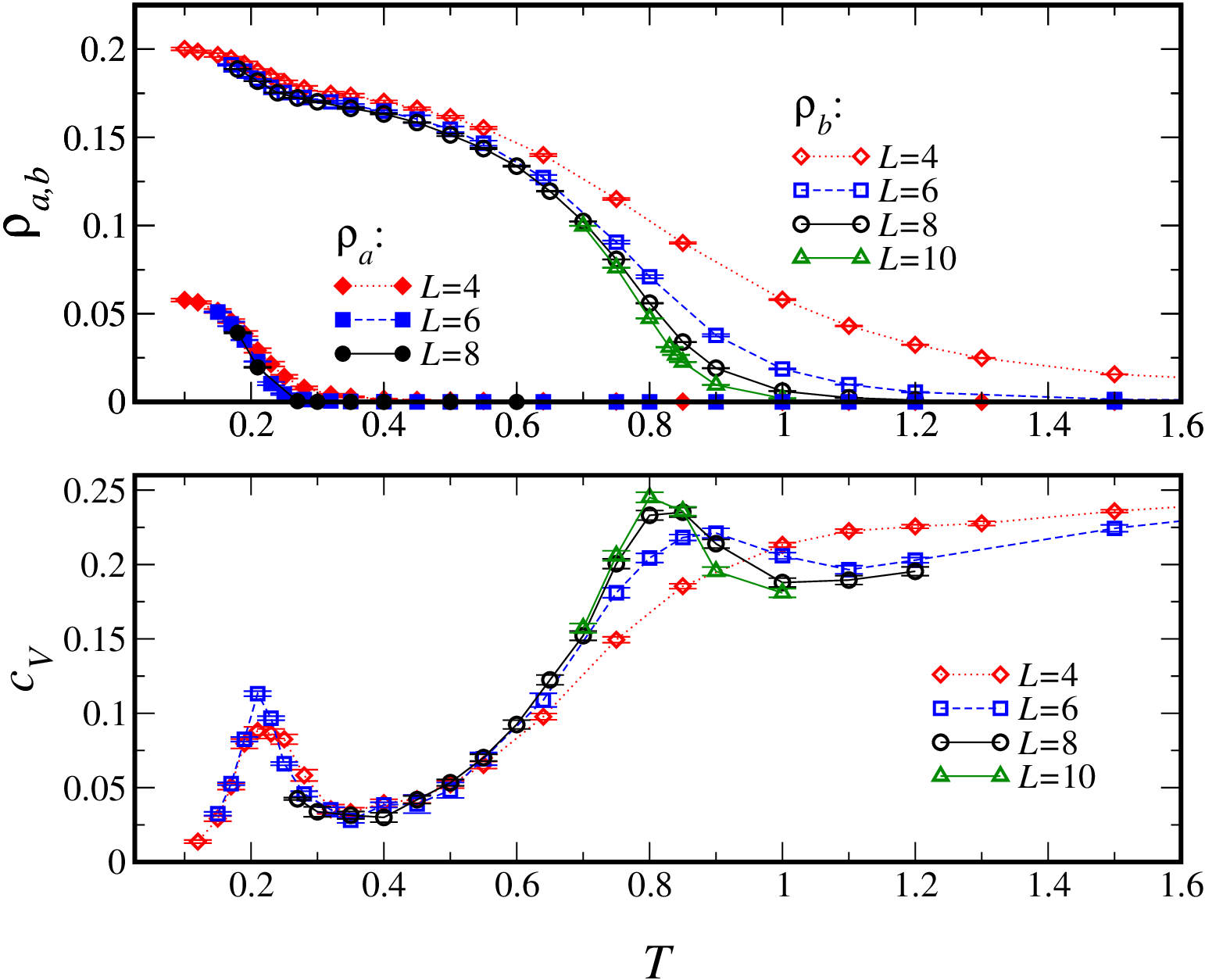}
\caption{Superfluid densities $\rho_a$ and $\rho_b$ 
and specific heat as functions of $T$ for non-symmetric case at point
$B$ indicated in Fig. \ref{444T015phase}(a). Here the low temperature phase is the VI-SF phase where both 
$a$-boson and $b$-boson condense. 
}
\label{VISF}
\end{figure}

In Fig. \ref{VISF} we show $\rho_{a,b}$ and $c_V$ for the system at the point $B$ where
there are two transitions.  As the temperature is lowered, the system first 
undergoes a transition at $T=0.814\pm0.002$ from a
normal-solid into the V-SF(B)-phase, in which the $b$-bosons condense. 
As the temperature is further lowered,
the system undergoes a second phase transition at $T=0.23\pm0.01$ within the supersolid state from
the V-SF(B) phase to the VI-SF-phase 
where both the $a$- and $b-$bosons condense. The critical temperatures have again been estimated by a scaling analysis similar to the symmetric case.
Note that below the lower transition point, $\rho_b$ further increases, due to the attractive interaction with the $a$-bosons which effectively increase the chemical potential for the $b$-bosons and hence their number. We have calculated $\rho_{\pm}$ and have found 
that $\rho_+$ and $\rho_-$ are almost the same,
so that the correlations are very small.

\begin{figure}[tbp]
     \centering
    \includegraphics[width=\columnwidth]{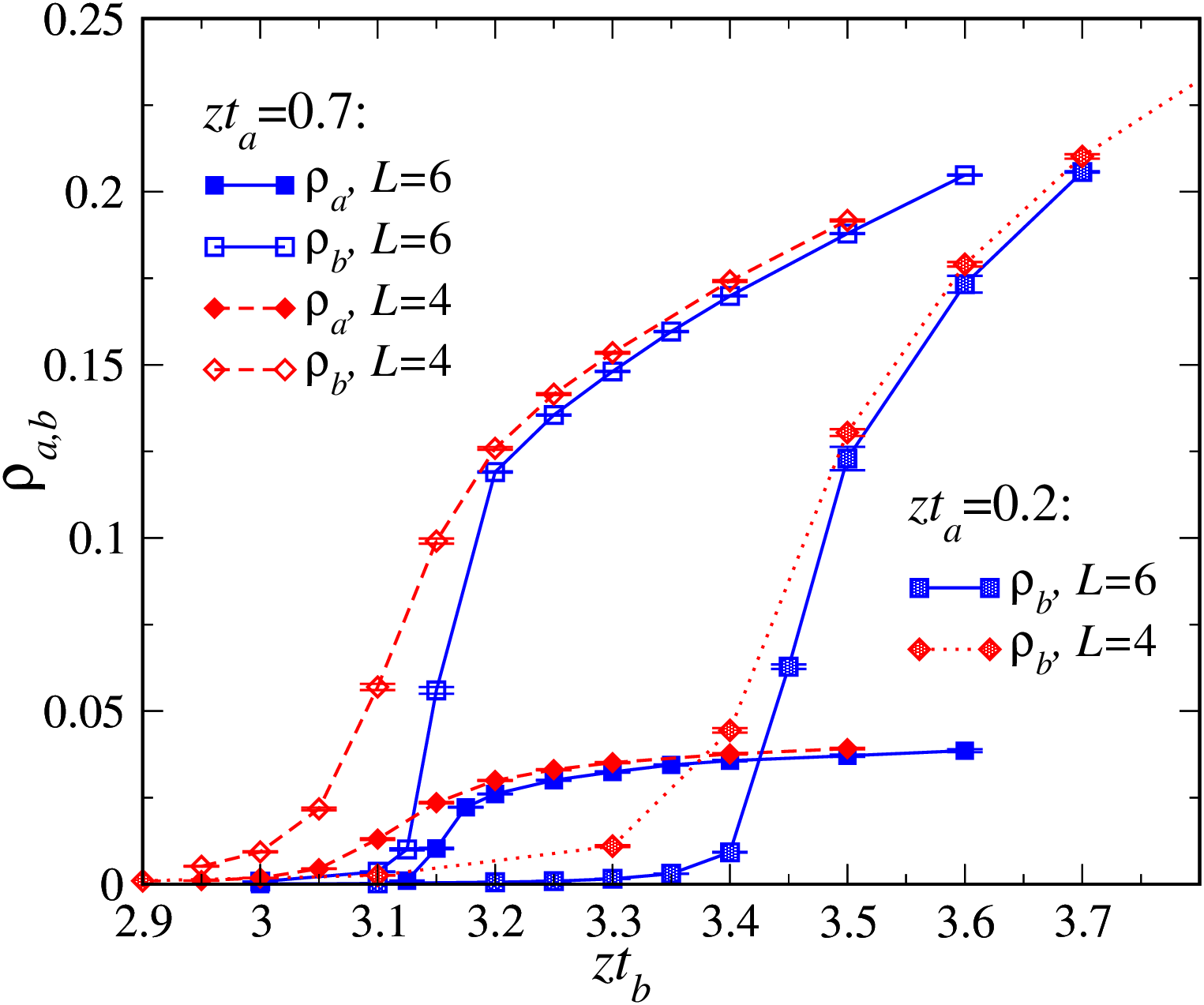}
    \caption{Superfluid densities $\rho_a$ and $\rho_b$ as functions of $zt_b$ for $zt_a=0.2$ and $zt_a=0.7$
along the lines $C$ and $D$ of Fig. \ref{444T015phase} at $T=0.15$. Note that
$\rho_a=0$ for $zt_a=0.2$.
}
    \label{QPTzta07zta02}
\end{figure}

\begin{figure}[tbp]
     \centering
    \includegraphics[width=\columnwidth]{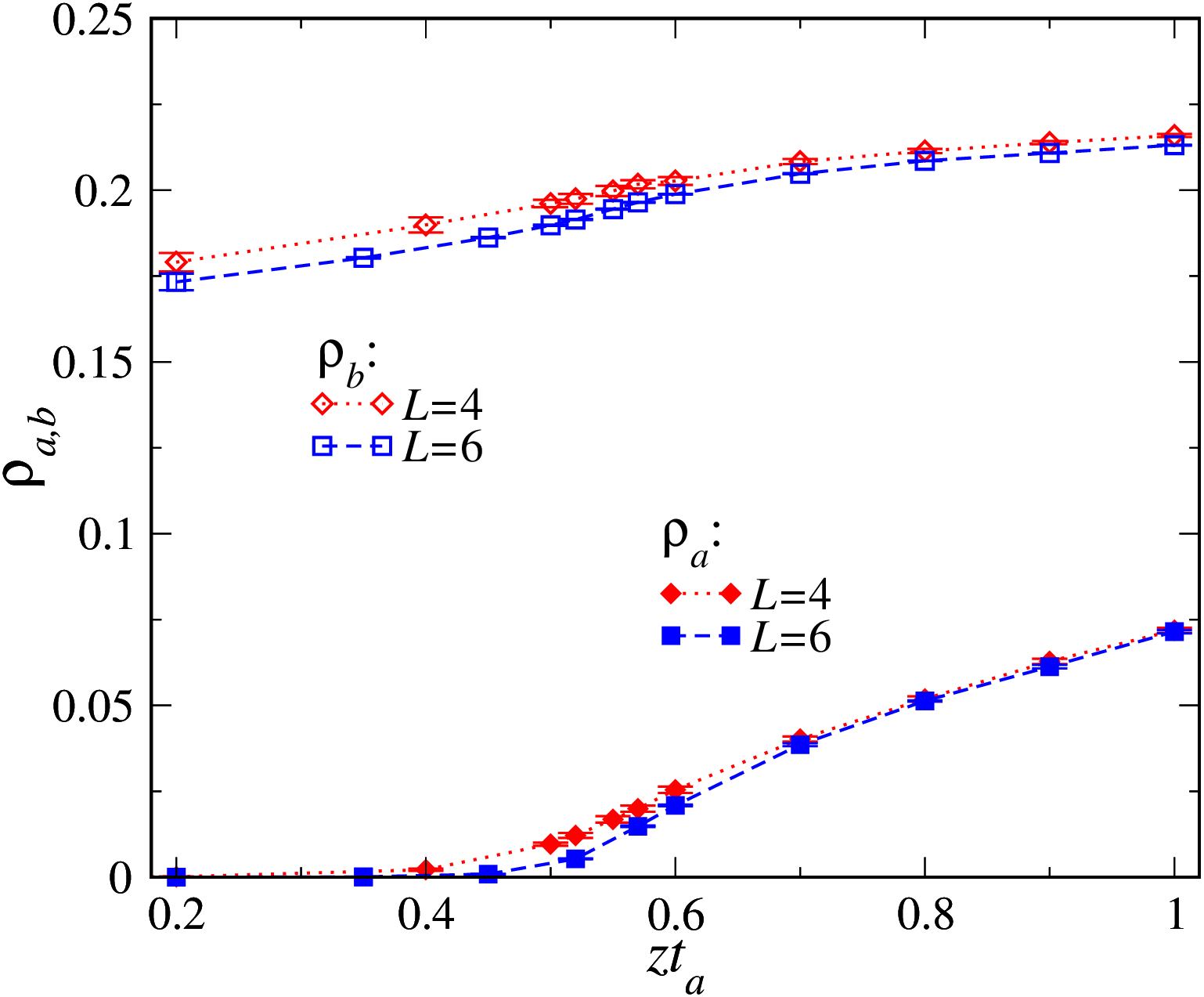}
    \caption{Superfluid densities as functions of $zt_a$ for non-symmetric case along line $A$ in 
Fig. \ref{444T015phase}(a) at $T=0.15$. At this temperature, $\rho_b > 0$, and the transition
is between the two supersolid phases V-SF(B) and VI-SF.
}
    \label{QPTztb36}
\end{figure}

We now discuss the phase transitions along the lines $C$-$F$ of Fig. 6(a) in more detail, at a temperature $T=0.15$. 
There are three different phase-transitions:
\begin{enumerate}
\item a transition between the insulating state and the V-SF(B)
phase and along the line $C$ in Fig. \ref{444T015phase}(a). For fixed $zt_a=0.2$,
we pass through a critical value where $\rho_b$ becomes 
non-zero while  $\rho_a$ remains zero (see Fig. \ref{QPTzta07zta02}). We estimate the critical value 
$zt_b= 3.44 \pm 0.02$ using the scaling analysis. 
\item a transition between the insulating state and the VI-SF phase appears along the lines $D$ and $E$ in 
Fig. \ref{444T015phase}(a). 
For fixed $zt_a=0.7$, 
the superfluid densities for both $a-$ and $b$-bosons are zero at small values of 
$zt_b$, and become finite above a critical value, which is the same for the two types of bosons, as we can see from Fig. 
\ref{QPTzta07zta02}.
The critical value of $zt_b$ can be estimated using a scaling analysis for different 
system sizes up to $L=6$ which gives the critical values of $zt_b=3.16\pm 0.02$ from 
for the line $D$ and $zt_b=2.87\pm 0.02$ 
for line $E$.
\item a transition is between two supersolid phases along line $F$ in Fig. \ref{444T015phase}(a).
Along this line $zt_b=3.6$  and $\rho_b$ is always finite.  As $zt_a$ increases $\rho_a$ is zero up to a critical value  $zt_a=0.55\pm 0.01$.  (see Fig. \ref{QPTztb36}). 
\end{enumerate}

\section{Conclusions}
Our quantum Monte Carlo simulations of a two-species bosonic Hubbard model of a supersolid show a phase diagram qualitatively consistent with previous mean field results.\cite{PRBDMF} The attractive interaction between a vacancy and interstitial  may facilitate the superfluidity in a bosonic solid, even when single vacancies or interstitials are gapped. Quantum fluctuations which are ignored in the mean-field calculations stabilize the superfluid phase over a larger parameter regime. Unlike the modified spin wave calculations\cite{Zhuang} which finds first order phase transitions at finite temperatures, the quantum Monte Carlo calculations show consistency with the expected scaling behavior at second order phase transitions in the $U(1)$ universality class, for temperatures above $T=0.10$.

\acknowledgements

The QMC calculations have been carried out on the clusters HPCPOWER of HKU's Computer Centre and Hreidar of ETH Z\"urich. We wish to thank Michael Ma and Xi Dai for many useful discussions. 
The work was partly supported by Hong Kong's RGC grant.

\end{document}